\documentclass[final,3p,times,authoryear]{elsarticle}

\makeatletter
\def\ps@pprintTitle{%
 \let\@oddhead\@empty
 \let\@evenhead\@empty
 \def\@oddfoot{{ Accepted:  {\it Infectious Disease Modeling}}\hfill{\it December 10, 2025}}
 \let\@evenfoot\@oddfoot}
\makeatother

\usepackage{amssymb}
\usepackage{amsmath} 
\usepackage{color}
\usepackage{subfigure}

\usepackage{setspace}

\usepackage[authormarkup=none]{changes}
\definechangesauthor[name=LRast, color=red]{lrast}
\definechangesauthor[name=LR2, color=olive]{LR2}
\makeatletter
\@namedef{Changes@AuthorColor}{red}
\colorlet{Changes@Color}{red}
\makeatother

\journal{Infectious Disease Modelling}

\begin{document}

\begin{frontmatter}



\title{Determining disease attributes from epidemic trajectories}



\author[add2]{Mark P. Rast\corref{cor1}}
\affiliation[add2]{organization={Department of Astrophysical and Planetary Sciences, Laboratory for Atmospheric and Space Physics, University of Colorado},
           city={Boulder},
           postcode={80309}, 
           state={CO},
           country={USA}}
\author{Luke I. Rast}
\cortext[cor1]{Corresponding author:  mark.rast@colorado.edu}

\begin{abstract}
Effective public health decisions require early reliable inference of infectious disease properties. 
In this paper we assess the ability to infer infectious disease attributes from population-level stochastic epidemic trajectories.  In particular, we construct stochastic Kermack-McKendrick model trajectories, sample them with and without observational error, and evaluate inversions for the population mean infectiousness as a function of time since infection, the infection duration distribution, and its complementary cumulative distribution, the infection survival distribution.
Based on an integro-differential equation formulation we employ a natural regression approach to fit the corresponding integral kernels and show that these disease attributes are recoverable from both multi-trajectory inversions and regularized single trajectory inversions.  Moreover, we demonstrate that the infection duration distribution (or alternatively the infection survival distribution) and population mean infectiousness kernel recovered can be used to solve for the individual infectiousness profile, the infectiousness of an individual over the duration of their infection, assuming that individual infectiousness profiles are self-similar across individuals over the infection duration period. The work suggests that, aggressive monitoring of the stochastic evolution of a novel infectious disease outbreak in a single local well-mixed population can allow determination of the underlying disease attributes that characterize its spread.

\end{abstract}

\begin{keyword}
infectious disease epidemiology
\sep stochastic SIR model
\sep inverse problem
\sep disease attribute kernels
\sep Poisson generalized linear model

\end{keyword}
\end{frontmatter}


\section{Introduction}\label{sec:intro}

We examine the ability to infer infectious disease attributes from stochastic epidemic trajectories.  Our work is motivated by the need for reliable inference of the underlying disease infectiousness and infection duration for effective public-health decisions~\citep[e.g.,][]{10.3389/fpubh.2024.1408193}, particularly those regarding recommendations for isolation or quarantine.  While these can be obtained by developing a robust understanding of the pathogen, in a public health emergency it may be necessary to recover them from models~\citep[e.g., ][]{bershteyn2022, CORI2024100784} based on the progression of the infectious disease outbreak in population level data. Here we focus on the inverse problem, given an epidemic model, how well can we determine the disease attributes from stochastic trajectories?   

Many previous authors have examined the structural and practical identifiability of deterministic Susceptible-Infectious-Recovered (SIR) and Susceptible-Exposed-Infectious-Recovered (SEIR) model rate parameters~\citep[see e.g.,][and references therein]{capaldi2012, tuncer2018, PhysRevResearch.2.043028, melikechi2022, sauer2022, cunniffe2023}.  A variety of inversion  and iterative forward-modeling approaches have been investigated including, ordinary and generalized least-squares estimation~\citep[e.g.,][]{capaldi2012, melikechi2022}, iterative minimization of ensemble or Poisson Kalman filter forecasts~\citep[e.g.,][]{PhysRevResearch.2.043028, sauer2022}, and  Markov chain Monte Carlo (MCMC) methods~\citep[e.g.,][]{talawar2016,sauer2022}.  These efforts have shown that the rate parameters are structurally identifiable in the SIR model and locally structurally identifiable in the SEIR model when either prevalence data or cumulative incidence data is used.  In the presence of Gaussian measurement error, they are practically identifiable from prevalence data in SIR but not in SEIR models, and are not practically identifiable for either model from cumulative incidence data~\citep{tuncer2018}.  Importantly,  identifiability, when possible, requires that the outbreak peak be sampled. This is true of both fully deterministic~\citep{tuncer2018} models and those with Poisson noise~\citep{sauer2022}.  
In this paper, we build on this line of work in two dimensions.
First, rather than fitting rate coefficients for infection and recovery processes, we focus on fitting integral kernels that describe how the rates of these process evolve within each individual host as a function of time since they were initially infected.
Second, we focus on (practical) identification in the context of a large epidemic, one that is potentially composed of multiple well-mixed populations. 
Instead of a limited population trajectory length, we explore the limitations imposed by small numbers of full length trajectory samples.

Ordinary or generalized least square inversion of deterministic SIR trajectories for the rate parameters that change with time has also been investigated~\citep{marinov2022}, and calibration of SIR rate coefficients by iterative non-linear least squares fitting to COVID -19 epidemic trajectories has been attempted~\citep{2020PhyD..41332674C}.   Further, use of cumulative-incidence time series to determine the force of infection~\citep[e.g.,][and references therein]{hens2010} or the population mean infectiousness kernel has been explored~\citep{pijpers2021}. 
In the latter work, regularized Fourier deconvolution was evaluated as an inversion technique using synthetic cumulative-incidence time series before being applied to early publicly available Covid-19 hospitalization data to determine the epidemic's effective reproduction number.
Our approach builds on this line of work, both by fitting additional individual-scale disease attributes, and by extending it to fully stochastic trajectories, where we find a natural correspondence to a GLM regression.

We construct stochastic SIR trajectories assuming random-walk contacts, specified individual-infectiousness profiles (individual infectiousness as a function of time since infection), and specified infection-duration probability distributions.   
We evaluate inversions of these stochastic trajectories, coarsely sampled with and without observational error, for the full suite of SIR integral kernels:  the population mean infectiousness, the infection duration distribution, and its complementary cumulative distribution, the infection survival distribution. 
We also use these recovered kernels to determine a characteristic profile for an individual's infectiousness as a function of time, assuming self-similarity for the disease progression.
The recovered profiles and distributions are directly comparable to those used in construction of the stochastic trajectories.  
We assess the the reliability of the inversion method employed for both multi and single trajectory inversions in the presence of the stochastic noise inherent in the model, and evaluate the influence of observational error to gauge practical inversion feasibility.

We note that in the context of a larger epidemic we are asking here whether observations of the stochastic evolution of outbreaks in local well-mixed populations allow determination of the underlying disease attributes, leaving aside questions about how the interactions between such local populations on larger spatial and temporal scales produce a full-scale epidemic.  Consistent with previous work, we use the term `epidemic' for these outbreaks in local well-mixed populations.
  
\section{Formulation for Integral Kernel Inversion}\label{sec:form} 

We employ the integro-differential framework for the general well-mixed closed-population model of~\cite{KM27} as developed by ~\cite{inaba2001} and~\cite{breda12}.  The susceptible $s$, infected $i$, and recovered $r$ population sizes evolve as,
\begin{equation}
{\frac{ds}{dt}}=-F(t)s(t) \ , \label{eq:KM1} 
\end{equation}
\begin{equation}
{\frac{di}{dt}}=F(t)s(t)-\int_0^t F(t - \tau)\,s(t - \tau)\,D(\tau)\,d\tau\ ,\label{eq:KM3a}  
\end{equation}
or equivalently 
\begin{equation}
{\frac{dr}{dt}}=\int_0^t F(t - \tau)\,s(t- \tau)\,D(\tau)\,d\tau \ , \label{eq:KM3b}
\end{equation} 
with $s+i+r=s_0$, the total population size, at all times.
Here, $\tau$ is the time post infection, and the force of infection,
\begin{equation}
F(t) = \int_0^t F(t - \tau)\,s(t - \tau)\,A(\tau)\,d\tau \ ,\label{eq:KM2}
\end{equation}
describes the the overall level of exposure to infection felt by individuals in the population.
It is the integral over the number infections occurring at previous times, $F(t - \tau)\,s(t - \tau)$, weighted by the {\it population mean infectiousness kernel} evaluated $\tau$ time units post infection $A(\tau)$.  The integral kernel $A(\tau)$ is thus the expected infectiousness of individuals infected $\tau$ time units in the past.

Similarly, the integral kernel $D(\tau)$ is the \textit{infection duration kernel}, the fraction of infected individuals who recover (become noninfectious for all future times) after $\tau$ units of time post-infection. 
This can be alternatively written~\citep{KM27} in terms of an {\it infection survival kernel}, $B(\tau)$, the fraction of individuals who remain infected $\tau$ time units after becoming infected, so that that the infected population evolves as 
\begin{equation}\label{eq:KM3c}
i(t)=\int_0^t F(t-\tau)\,s(t-\tau)\,B(\tau)\,d\tau\ . 
\end{equation}
In the discrete case, Section~\ref{sec:model} below, $D$ is the infection duration probability distribution with $B$ its complementary cumulative distribution.

We are also interested in recovering the characteristic \textit{individual infectiousness profile}, which we denote $I$, the infectiousness of an individual as a function of time within the infected period.
We take the individual infectiousness profile $I$ to be self-similar across individuals, meaning that the individual infectiousness profiles scale with the infection duration $T$, so that  $I\left(\tau/T\right)$ is identical for all $T$; the infectiousness of an individual progresses in the same way for all individuals with its progression rate determined by the total duration of the infection.  
Under this assumption,  
\begin{equation}\label{eq:I1}
    A(\tau)= \int_\tau^\infty I\left(\tau/T\right) D(T)\,d\,T\, , 
\end{equation}
where
the lower integral bound reflects that $I\left(t\right)=0$ for all $t>T$.
The population mean infectiousness kernel $A(\tau)$ is then the expectation over the individual infectiousness profiles given their duration.  It captures the average infectiousness with time after infection, reflecting both the individual infectiousness and infection duration.  The population size $s_0$ times the integral of $A(\tau)$ over all $\tau$ yields the basic reproduction number $R_0$, the average number of secondary infections caused by a single infected individual.  Thus $A$ and $I$ in this formulation, with dynamics independent of infection, include 
an underlying contact rate scaling~\citep[e.g.,][]{breda12}.

These equations are very similar to the classic rate-based SIR model~\citep[also from][]{KM27}, albeit with the rate parameters for infection and recovery processes replaced by the infectiousness and infection-duration kernels $A(\tau)$ and $D(\tau)$.  We note, however, that in the classic SIR model infected and infectious have the same meaning, $I\left(\tau/T\right)=1$.  Here we take the infectiousness to vary within the infected period, even allowing non-infectious portions within that period.  In this context, the recovered population $r$ consists of those individuals who are never possibly infectious in the future, and in the classic SIR sense never possibly reinfected. 

\subsection{Discrete stochastic model}\label{sec:model}

As emphasized by \cite{2021PNAS..11806332D}, epidemic time series are not continuous.
The processes underlying them are inherently discrete and stochastic, and measurements of the populatio are usually taken at  time-intervals longer than the time between individual infections or recoveries.
With this in mind, we can derive discrete stochastic versions of Equations~(\ref{eq:KM1}),~(\ref{eq:KM3b}), and~(\ref{eq:KM3c}) for the observed changes in the number of susceptible, recovered, and infected individuals (see \ref{sec:appendix:stochastic_model_derivation}).
In expectation, they become
\begin{equation}
\mathbb{E}\left[\Delta s_k\vert\Delta s_0\,.\,.\,.\,\Delta s_{k-1}\right] =\ s_k \sum_{n=1}^{k} \Delta s_{k -n} A_{n} \label{eq:KM5}
\end{equation}
\begin{equation}
-\,\mathbb{E}\left[\Delta r_k\vert\Delta s_0\,.\,.\,.\,\Delta s_{k-1}\right] =\sum_{n=1}^{k} \Delta s_{k -n} D_{n} \label{eq:KM6}
\end{equation}
\begin{equation}
\mathbb{E}\left[i_k\vert\Delta s_0\,.\,.\,.\,\Delta s_{k-1}\right] =\sum_{n=1}^{k} \Delta s_{k -n} B_{n} \label{eq:KM7} 
\end{equation}
where $s_k$, $i_k$, $r_k$, $A_n$, $D_n$, and $B_n$ are the discrete versions of their counterparts in the integral equations.
We show in~\ref{sec:appendix:stochastic_model_derivation} that the distribution of $\Delta s_k$ is approximately Poisson, while the distributions of $\Delta r_k$ and $i_k$ can be approximated as Normal.
In this paper, we will take the sampling interval $\Delta t_k$, over which changes $\Delta s_k$ and $\Delta r_k$ occur, to be constant, but uniform temporal sampling is not essential in this formulation. 

These equations resemble autoregressive time series, with the caveat that $\Delta s_k$ depends on the current $s_k$ value.
They are equivalent to their continuous counterparts for infinitesimally small increments and large populations, but conceptually they precede the continuous form~\citep{KM27} and capture the discrete and stochastic nature of the underlying epidemic processes.
At time $t_k$, Equation~(\ref{eq:KM5}) relates the expected incidence of infection $\Delta s_k$ to the current susceptible population $s_k$, all past incidences of infection $\Delta s_{k-n}$, and $A_{n}$, the population mean infectiousness after a time interval $n\Delta t$.
Similarly, from Equation~(\ref{eq:KM6}), the expected number of recoveries in the time interval $\Delta t_k$ depends on the all past incidences $\Delta s_{k-n}$ and $D_{n}$, the probability with which individuals recover after a time interval of $n\Delta t$.
Likewise, Equation~(\ref{eq:KM7}), indicates that the expected number of infected individuals at time $t_k$ depends all past incidences $\Delta s_{k-n}$ and $B_{n}$, the fraction of those who remain infectious after a time interval of $n\Delta t$.

Equation~(\ref{eq:I1}) can also be written in a discretized form (\ref{sec:appendix:stochastic_model_derivation}) as,
 \begin{equation}\label{eq:I2_main}
A_n=\sum_{k=n}^{N-1}\  {{n\,T_{\rm max}}\over{k^2}}\ D\!\left({{n}\over{k}}\,T_{\rm max}\right)I\!\left({k\over{N}}\right)\ ,  
\end{equation}
where $N$ is the number of composite-quadrature points, chosen so that time is discretely sampled at the same constant interval that defines the epidemic time series ($\tau$ in Equation~(\ref{eq:I1}) equals $n\,T_{\rm max}/N$), and $T_{\rm max}$ is the maximum infectious duration, so that $D(T>T_{\rm max})=0$.  After recovery of $A_n$ and $D_n$ by the inversion of Equations~(\ref{eq:KM5}) and~(\ref{eq:KM6}) (or~\ref{eq:KM7}), Equation~(\ref{eq:I2_main}) allows the self-similar individual infectiousness profile $I(n/N)$ to be estimated. 

We note that the discrete kernels $A_n$, $D_n$, and $B_n$ in Equations~(\ref{eq:KM5}) - (\ref{eq:I2_main}) are binned temporal-averages over a time interval $\Delta t_n$.  Given stochastic time series $s_k$, $i_k$, and $r_k$, we aim to recover $A_n$, $D_n$, and $B_n$ by regression, and $I_n$ via Equation~(\ref{eq:I2_main}) from them.  
For the remainder of this paper we take the discrete kernels to be defined on the same observational grid as the time series and can thus drop the subscripts $n$ and $k$. 

\subsection{Observational error}\label{sec:obs}

In addition to the intrinsic stochastic noise in $\Delta s$ and $\Delta r$ (modeled above), which reflects the development of the epidemic itself, observations may also contain additional noise, in the form observational error.
The observational strategy we have in mind is one in which there is frequent self-testing by the population, who then report when they become infected (first positive test result) or recover (first negative test result).
The reported $\Delta s$'s and $\Delta r$'s, along with knowledge of the total population size $s_0$, are then used to construct all the quantities needed for recovery of the kernels.

As a result of this data collection strategy, only a fraction of the true $\Delta s$ and $\Delta r$ changes may be captured and false positives may occur as well;  individuals may fail to report changes in their infection status or give false-positive reports of those changes.  We take the former to occur with a probability of $1-p_{\textrm{T}}$  and the later with a probability of $p_{\textrm{F}}$.  For simplicity, when we evaluate their effect on the kernel inversions, we will take these errors to occur independently at each time step, ignoring any possible error correlations between variables or time steps.  

Under these reporting model assumptions, we can construct observed epidemic trajectories from the true stochastic trajectories (as constructed in Section~\ref{sec:model2} below).  While the true time series, 
{\it s}, {\it i}, and {\it r}, account for each individual infection and recovery, they are observationally subsampled over larger $\Delta t_{\rm obs}$.
When accounting for reporting error, over that observational $\Delta t_{\rm obs}\,$,
\begin{equation}\label{e1}
    \Delta s_{\rm obs} = X(\Delta s, p_{\textrm{T}})+X(s, p_{\textrm{F}})\, , 
\end{equation}
\begin{equation}\label{e2}
    \Delta r_{\rm obs} = X(\Delta r, p_{\textrm{T}})+X(i, p_{\textrm{F}})\, ,
\end{equation}
where $X\sim {\rm Bin}(n, p)$; the random variates follow the binomial distribution with $n$ trials and a probability $p$ of success.
The observed/reported changes in the number of susceptible individuals (the number newly infected) or the number of recovered individuals thus captures errors due to both incomplete reporting (first terms on the right hand sides of Equations~(\ref{e1}) and~(\ref{e2})) and false positive reports (second terms on the right hand sides of Equations~(\ref{e1}) and~(\ref{e2})).

For the $A$- and $B$-kernel regressions (Equations~(\ref{eq:KM5}) and~(\ref{eq:KM7})) $s_{\rm obs}$ and $i_{\rm obs}$ are also required.  Under the reporting scenario outlined, these must be obtained by accumulation of the observed $\Delta s$'s and $\Delta r$'s.  At each time step, summing over all previous reported incidences,
\begin{equation}\label{e3}
    s_{\rm obs} = \sum\Delta s_{\rm obs}\, ,
\end{equation}
\begin{equation}\label{e4}
    r_{\rm obs} = \sum\Delta r_{\rm obs}\, ,
\end{equation}
and
\begin{equation}\label{e5}
    i_{\rm obs} = s_0- s_{\rm obs} - r_{\rm obs}\, .
\end{equation}
With reporting errors in $\Delta s_{\rm obs}$ and $\Delta r_{\rm obs}\,$, $\,i_{\rm obs}$ can become negative, an unphysical outcome, and we take $i_{\rm obs}=0$ whenever that occurs during accumulation. 

\section{Stochastic Trajectory Differences from Rate-based SIR}\label{sec:model2}

\subsection{Epidemic kernels employed}\label{sec:kernels}

\begin{figure}[t!]
\vskip -1cm
\centerline{\includegraphics[scale=0.5]{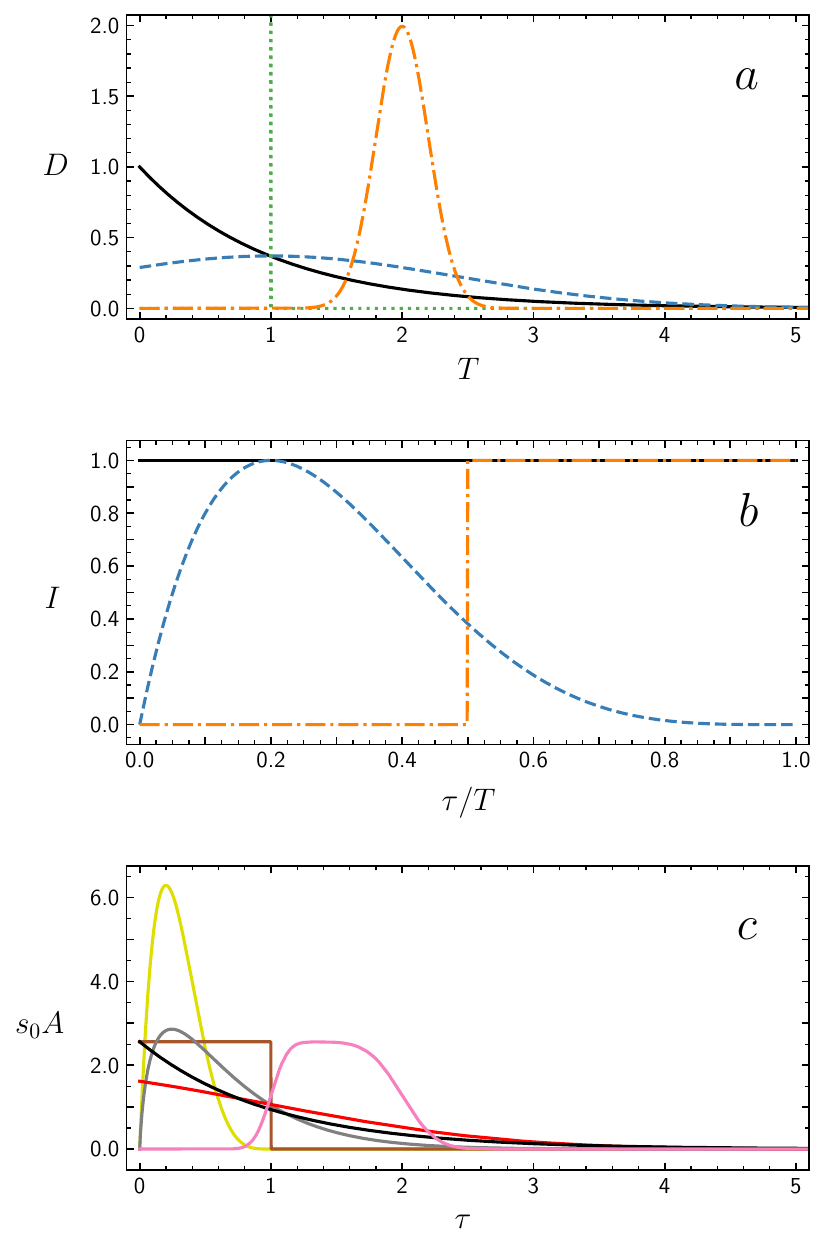}}
\caption{Disease attributes kernels used to construct stochastic trajectories.
In ($a$), the four infectiousness duration distributions $D(T)$ employed (exponential {\it solid black}, normally distributed with peak at $\tau_{\!{}_D}$ {\it dashed blue}, $\delta$-distributed with value $\tau_{\!{}_D}$ {\it dotted green}, and normally distributed with a mean of $2\tau_{\!{}_D}$ {\it dash-dot orange}).  In ($b$),  the three self-similar infectiousness profiles employed (constant over the infected period {\it solid black}, beta distribution shaped {\it dashed blue}, and taking the form of a Heaviside step function {\it dash-dot orange}).  In ($c$), six population mean infectiousness profiles that result (see text for more detailed decription):  EC (exponential-constant, {\it black}), N1C (normal-constant, {\it red}), N1B (normal-beta, {\it grey}), N2H (offset normal-step function {\it pink}), DC ($\delta$ function-constant, {\it brown}), and DB ($\delta$ function - beta, {\it gold}).
}
\label{fig:disease_attributes}
\end{figure}
  
The fully-resolved error-free stochastic trajectories {\it s}, {\it i}, and {\it r} we construct are based on random-walk contact statistics and specified infection duration distributions and individual infectiousness profiles.  Motivated by the desire to examine inversion reliability over a range of distinct kernel shapes,
we employ four different duration distributions and three self-similar individual infectiousness profiles in their construction.
These are plotted in Figure~\ref{fig:disease_attributes}$a$ and $b$.  Of the twelve possible infection duration distribution/individual infectiousness profile combinations, we present, in this paper, results using six.  
The population-average infectiousness kernels $A$ corresponding to those six combinations are shown in Figure~\ref{fig:disease_attributes}$c$.   

In detail, the infectiousness duration distributions examined cover a range widths and offsets:  $D(T)$  is taken to be either exponentially distributed with a characteristic decay time of $\tau_{\!{}_D}$ (Figure~\ref{fig:disease_attributes}$a$ {\it solid black} line style), the positive part of a normal distribution with peak location and standard deviation equal to $\tau_{\!{}_D}$ (Figure~\ref{fig:disease_attributes}$a$ {\it dashed blue} line style), $\delta$-distributed with value $\tau_{\!{}_D}$ (Figure~\ref{fig:disease_attributes}$a$ {\it dotted green} line style), or normally distributed with a mean of $2\tau_{\!{}_D}$ and standard deviation $0.2\tau_{\!{}_D}$ (Figure~\ref{fig:disease_attributes}$a$ {\it dash-dot orange} line style).  
With these definitions we have adopted a universal scaling for time, taking time to always be measured in units of the mean infectious duration of the exponential distribution $\tau_{\!{}_D}$.  
We note, that the exponential duration distribution with mean infectious duration $\tau_{\!{}_D}$, when combined with constant individual infectiousness over the infected period, yields the stochastic version of the classic rate based SIR model (Section~\ref{sec:model2} below).  The other cases may or may not have the same mean infection duration, but for comparison, we scale time by $\tau_{\!{}_D}$ in all plots.

The self-similar individual infectiousness profiles $I\left(\tau/T\right)$ we employ are constant, advanced, or  delayed:  constant over the infected period (Figure~\ref{fig:disease_attributes}$b$ {\it solid black} line style), beta distribution~\citep[e.g.,][]{casella2002} shaped with $\alpha=2$ and $\beta =5$ (Figure~\ref{fig:disease_attributes}$b$ {\it dashed blue} linestyle), or taking the form of a Heaviside step function at $\tau/T=0.5$ (Figure~\ref{fig:disease_attributes}$b$ {\it dash-dot orange} linestyle) respectively. 
The infectiousness profiles in Figure~\ref{fig:disease_attributes}$b$ are plotted with unit maxima, but,
importantly, since the duration distribution is normalized, the magnitude of $I$ can be scaled so that the epidemics all share the same underlying basic reproduction number $R_0$ via the integral over the population average infectiousness $A$ as determined from $D$ and the scaled $I$ (Equation~\ref{eq:I1}).

Those $A$ kernels for the combinations we present in this paper are shown in Figure~\ref{fig:disease_attributes}$c$, and we use the following short-hand notation to identify them:   EC, exponential duration distribution with constant infectiousness ({\it black}), N1C, normal ($\mu=1.0$, $\sigma=1.0$) duration distribution with constant infectiousness ({\it red}), N1B, normal ($\mu=1.0$, $\sigma=1.0$) duration distribution with beta-distribution shaped infectiousness ({\it grey}), N2H, offset normal ($\mu=2.0$, $\sigma=0.2$) duration distribution with Heaviside step function infectiousness ({\it pink}),
DC, $\delta$-distributed duration with constant infectiousness ({\it brown}), and DB, $\delta$-distributed duration with beta-distribution shaped infectiousness ({\it gold}).

\subsection{Trajectory properties}\label{sec:model2}

With these kernels we simulate stochastic epidemic trajectories as~\cite{Gillespie77}, accounting for each individual infection and recovery.  
We take two essential parameters that characterize the stochastic SIR trajectories from a reference direct-numerical-simulation of a two-dimensional population of noninteracting random walkers~\citep{2022PhRvE.105a4103R}:  the basic reproduction number $R_0$ and the characteristic global waiting time (time between successive binary contacts in the population) $\lambda$.
In that simulation, the individual reproduction number is measured to be Poisson distributed with mean $R_0 = 2.559$ and the global waiting time (though not the individual waiting time) is exponentially distributed with rate parameter $\lambda= R_0\,\rho_{\!{}_N}  /2\approx327.6$, where $\rho_{\!{}_N}$ is the number density of random walkers.  

With the basic reproduction number and $\lambda$ fixed, we construct epidemic trajectories for all six kernel combinations of Section~\ref{sec:kernels} by $a$) starting with an single infected individual, $b$) successively sampling the exponential global waiting-time distribution and stepping forward in time to the next contact, and $c$) taking the probability of a new infection at that contact time to be $2\,s\,i\,I/s_0^{\,2}$, so that, upon contact, infection occurs when one walker is infectious and the other is susceptible. The individual infectiousness $I$ at contact is chosen randomly from the current infectious members of the population based on the time since their infection, the duration of their infectious period ($T$ as sampled from $D$ upon infection), and the self-similar infectiousness profile $I\left(\tau/T\right)$ being employed.  No account is made for the elevated probability of repeat encounters over times shorter than ballistic mean-free-collision time~\citep{2022PhRvE.105a4103R}.  

\begin{figure}[t!]
\vskip -2.0cm
\centerline{\includegraphics[scale=0.6]{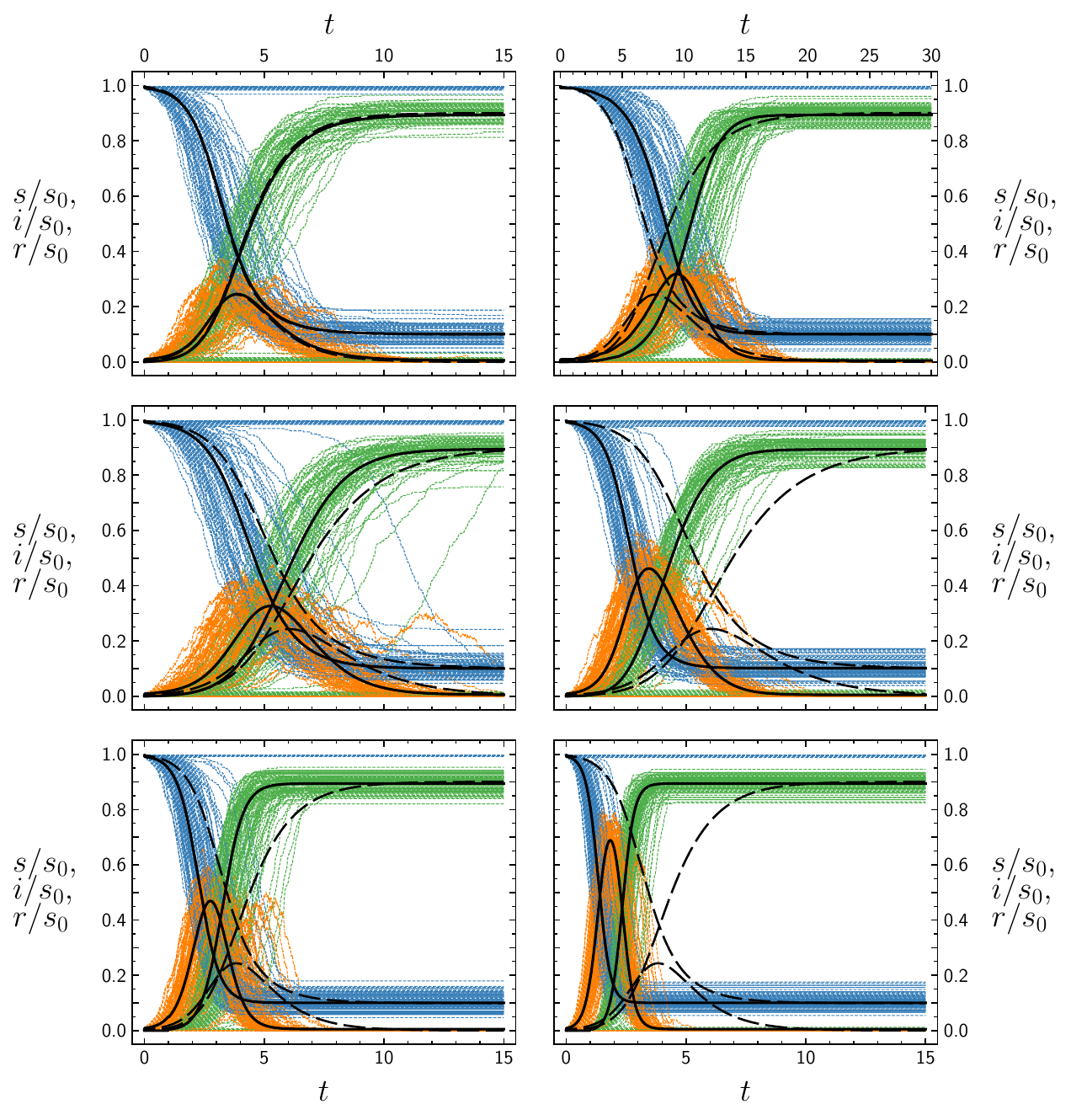}}
\caption{Rate-based SIR models fail to capture epidemic time-scales and the distribution of stochastic trajectories.  One hundred stochastic trajectories of $s$ ({\it blue}), $i$ ({\it orange}), and $r$ ({\it green}), normalized by $s_0$, are plotted for six epidemics all sharing the same $R_0$:  EC ({\it top} row, {\it left}), N1C ({\it middle} row, {\it left}), DC ({\it bottom} row, {\it left}), N2H ({\it top} row, {\it right}), N1B ({\it middle} row, {\it right}), DB ({\it bottom} row, {\it right}).
Over-plotted is the constant rate SIR solution ({\it dashed black} curves), taking the recovery rate to be the inverse of the mean infection duration and the infection rate the integral over the infectiousness profile.  Also plotted ({\it solid black} curves) are solutions to the deterministic~\cite{KM27} integro-differential Equations~(\ref{eq:KM1})~-~(\ref{eq:KM2}) with  same attribute kernels as the stochastic trajectories.  Note, only the {\it top right} panel has an extended time axis; time ranges from zero to $15\tau_{\!{}_D}$ in all other plots.
}
\label{fig:stochastic_trajectories}
\end{figure}

The resulting stochastic trajectories for each of the six kernels combinations (those for which $A$ is shown in Figure~\ref{fig:disease_attributes}$c$) are plotted in Figure~\ref{fig:stochastic_trajectories}, with the fraction of susceptible, infected, and recovered individuals shown for each epidemic trajectory in {\it blue}, {\it orange}, and {\it green} respectively.
Note that, due to stochasticity, epidemics may not spread if the early infections randomly fail to make sufficient contacts that result in further infections.
These failed epidemics appear as trajectories with $s/s_0$ plateauing with a value near 1 and $r/s_0=i/s_0$ plateauing with values near 0.  The number of failed epidemics (taken as those with final $s > 0.95\,s_0$) range in number from about 6\% of the DC epidemics to more than 40\% of the EC cases.

Plotted along with the trajectories in Figure~\ref{fig:stochastic_trajectories} are the constant rate SIR epidemic trajectories for $R_0=2.559$ ({\it dashed black} curves) and the deterministic trajectories obtained by direct integration of the~\cite{KM27} integro-differential equations (Equations~(\ref{eq:KM1})~-~(\ref{eq:KM2}) above, {\it solid black} curves) using the kernels specified for the stochastic cases.
The constant rate SIR solutions are computed using one over the mean duration as the recovery rate $\gamma$ and $R_0/s_0$ times integral over the scaled infectiousness profile as the infection rate $\beta$, so that $\beta s_0/\gamma = R_0$ and $ \int_0^\infty s_0\,A\,d\tau =R_0$ for all kernel combinations.

As expected~\citep[e.g.,][]{breda12}, while the constant rate SIR solutions capture the mean population endstates of the epidemics, their evolution in time agrees with the evolution of the stochastic trajectories (and with the deterministic integro-differential solution) only when $A$ is exponential.  In our examples, $A$ is exponential only when $D$ is exponentially distributed and $I$ is a constant over the duration of the infectious period (the EC case, {\it top left} panel of Figure~\ref{fig:stochastic_trajectories}).  In all other cases, the timescale for the stochastic epidemic development does not match rate based SIR solution. 

The epidemic timescale depends on the details of the kernels not just the characteristic rates.  
Epidemic trajectories generated with variable infectivity and/or non-exponential duration distributions differ from SIR  trajectories, even when they share the same $R_0$ and the same average infection duration.
A good example of this is the DC epidemic ({\it bottom} row {\it left} panel  Figure~\ref{fig:stochastic_trajectories}).  It shares the same infectiousness profile with the EC epidemic ({\it top} row {\it left} panel Figure~\ref{fig:stochastic_trajectories}), constant over the infectious period, and has the same mean infection duration, and thus the same rate based SIR trajectories.  However, the stochastic epidemic develops much more rapidly when the duration is $\delta$-distributed than exponentially distributed.  In the DC case, because the infection duration the same for all individuals, the population mean infectiousness $A$ (Figure~\ref{fig:disease_attributes}$c$, {\it brown}) is more heavily weighted to short $\tau$ infections and the epidemic onset is more rapid.  The shortening of the epidemic time scale is even more pronounced for the DB epidemic ({\it bottom} row {\it right} panel Figure~\ref{fig:stochastic_trajectories}; Figure~\ref{fig:disease_attributes}$c$, {\it gold}) for which the infectiousness is also high at short times.  
While the mixed effects of the two kernels can be more subtle (e.g.; epidemic N1C), in general the characteristic epidemic trajectory timescale depends, not only on the number of new infections caused by each infected individual $R_0$ and the mean duration of the infection, as in rate based SIR, but also on the distribution of the infection duration and when during an individual's infected period they are infectious.   

For these reasons, direct integration of the integro-differential equations does a better job of capturing the characteristic evolution of the stochastic solutions ({\it solid black} curves in Figure~\ref{fig:stochastic_trajectories}). For example, omitting the failed epidemics which are not captured by the deterministic solution, the mean time at which $s=0.5s_0$ in the stochastic solutions is close to that  of the deterministic solution, differing by about 1.4 to 11\% for the six duration-distribution infectiousness-profile combinations shown.  Even so, the distribution of stochastic trajectories around the deterministic solution is typically not symmetric.  The development of the stochastic epidemics is most often steeper than predicted by the deterministic model. 

The trajectory differences between stochastic
epidemics for differing kernel choices allows inversion for the input kernels.
Some kernel combinations show larger trajectory variations around the deterministic solution than others.  For example case N1C ({\it middle} row {\it left} column of Figure~\ref{fig:stochastic_trajectories}) shows particularly large differences between individual trajectories.  These variations within a given case makes those inversions more challenging, particularly when they are based on a single trajectory. 

\section{Poisson GLM regression provides a natural inversion approach}\label{sec:invert}

Our goal is to use population level stochastic trajectories of $s$, $i$, or $r$ (such as those plotted in Figure~\ref{fig:stochastic_trajectories}) to recover the disease attributes, $D$, $B$, and $A$, and in turn $I$.
As it is unlikely, in any real application, that the population trajectories can
be sampled over time intervals short enough to capture each successive infection, as our stochastic simulations have, we coarsen the successful stochastic trajectories (those with final $s < 0.95\,s_0$) by sampling them at an interval of $0.05\tau_D$.  This is representative of what very aggressive monitoring of a population in the real world might be able to achieve, and also allows possible inversion for the most steeply rising kernels we hope to recover (for example the $A$ kernels plotted in {\it grey} and {\it gold} in Figure~\ref{fig:disease_attributes}$c$).
We show in Section~\ref{sec:glm1} that our method is robust to larger sampling intervals, with the main effect being a decrease in the time-resolution of the kernel fits, matching the sampling resolution.
With fixed sampling rate, we examine multi-trajectory inversions, both with (Section~\ref{sec:glm1b}) and without (Section~\ref{sec:glm1}) reporting error, and regularized individual trajectory solutions (Section~\ref{sec:glm2}).

\subsection{Regression method}\label{sec:glm0}

Because the contact statistics in the our stochastic model are Poisson, 
we construct a Poisson GLM to fit the epidemic kernels by regression; we model the stochastic changes in the number of susceptible, infectious, and recovered individuals in the population based on the previous infections (Equations~(\ref{eq:KM5}) $-$~(\ref{eq:KM7})) and find the maximum likelihood solution for the kernels.  
Thus, each data point (left-hand-side of Equations~(\ref{eq:KM5}) $-$~(\ref{eq:KM7})) is taken to be Poisson distributed (see~\ref{sec:appendix:stochastic_model_derivation}) with a mean that is the weighted sum of previous datapoints, earlier changes in the number of susceptible individuals (right-hand-side of Equations~(\ref{eq:KM5}) $-$~(\ref{eq:KM7})).
Since $\Delta s$, $\Delta r$, $i$, and $s$ in these equations are observable (possibly with observational error) from the stochastic population dynamics, we can fit for the weights $A$, $D$, and $B$ in the expressions for the mean of the Poisson distribution.
In the parlance of GLMs, this corresponds to a `link function' that is the identity function.
We constrain all elements of the recovered parameter vector to be positive, as must be the case for the ground-truth kernels. 

We also include the ability to regularize the solution if necessary.  As the
kernel vectors to be recovered are discrete versions of the smooth continuous
kernels that underlie the stochastic trajectories, when employing regularization, we favor parameter vectors with small changes between neighboring elements.
In order to enforce this smoothness, we add a term to the cost function
that penalizes large changes in adjacent values $\lambda_s \sum (A_n - A_{n-1})^2$ (similarly for $B_n$ and $D_n$).  This 
corresponds to a prior over neighboring parameter differences $(A_n - A_{n-1})$ that is Gaussian with mean zero and variance $1/ \lambda_s$.
As both the Poisson log-likelihood of the GLM and the squared-difference
smoothness constraint are convex in the parameter vector,
we can fit the model by convex optimization, which we do with the python package \texttt{cvxpy} (https://www.cvxpy.org/index.html), using primarily the convex solver \texttt{Clarabel} (or occasionally \texttt{SCS}) in that package (https://github.com/oxfordcontrol/Clarabel.jl).

Finally, once $A$ and $D$ are determined we use them to invert Equation~(\ref{eq:I2_main}) for the individual infectiousness $I$.  For this we employ the iterative sparse-matrix damped-least-squares (DLS) solver from the SciPy package \texttt{sparse.linalg.lsqr}~\citep[][]{10.1145/355984.355989}.  This algorithm allows specification of a damping parameter, here noted $\delta_{\rm lsqr}$.

\subsection{Multi-trajectory inversions}\label{sec:glm1}

\begin{figure}[t!]
\vskip -2cm
\centerline{\includegraphics[scale=0.65]{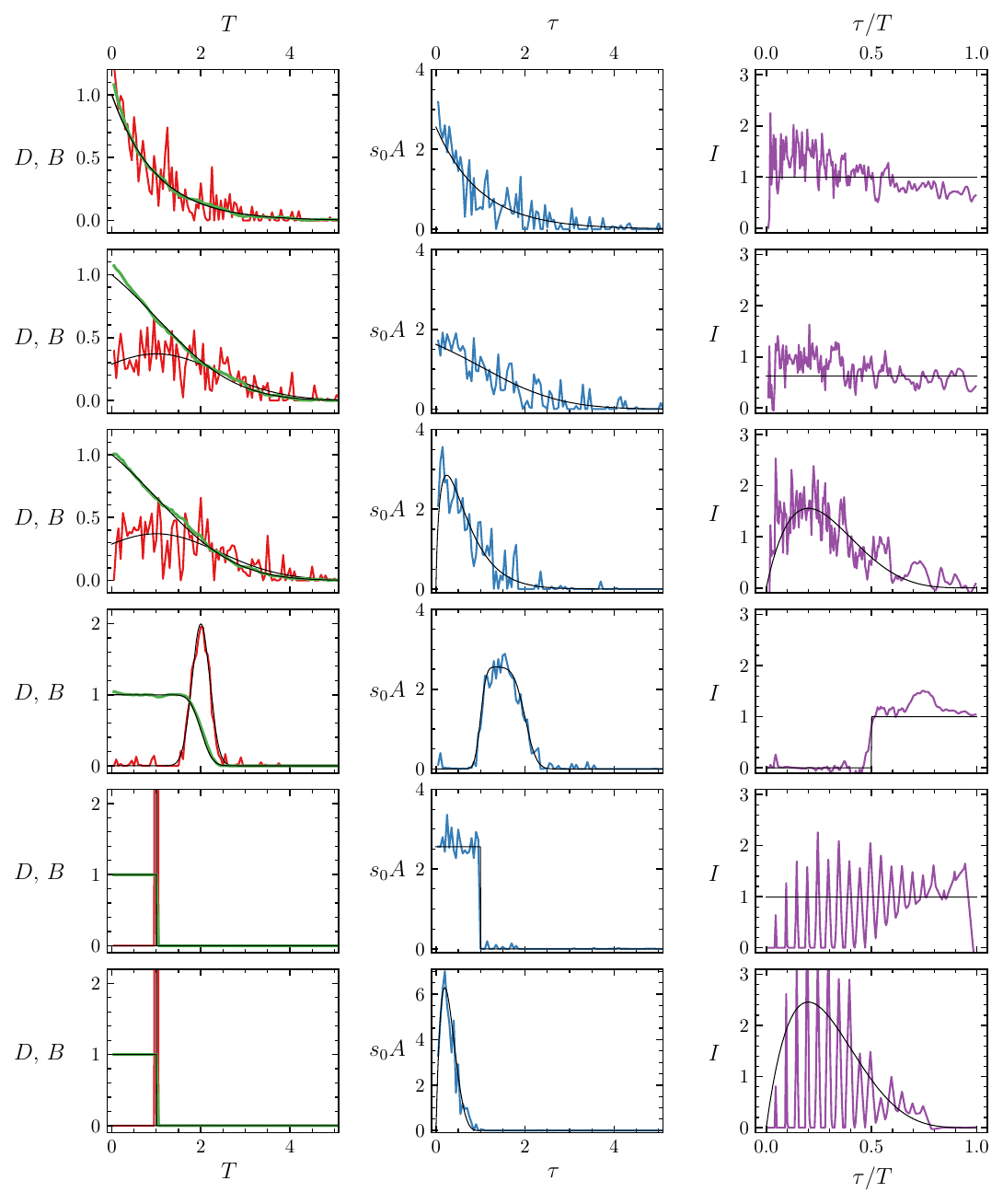}}
\caption{Multi-trajectory Poisson GLM inversions for the disease attribute kernels, $D$ ({\it red}) and $B$ ({\it green}) in the {\it first} column and $s_0A$ ({\it blue}) in the {\it middle} column, along with regularized least square inversions (using the recovered $D$ and $A$ kernels) for $I$ ({\it purple}) in the {\it third} column.  One-hundred stochastic trajectories (those plotted Figure~\ref{fig:stochastic_trajectories}) were employed, each constructed with the target kernels shown in {\it black}. Rows from {\it top} to {\it bottom}: EC, N1C, N1B, N2H, DC, and DB.}
\label{fig:multitrajectoryinversion}
\end{figure}

Figure~\ref{fig:multitrajectoryinversion} shows the results of Poisson GLM inversions for the duration distribution $D$, its complementary cumulative survival distribution $B$, and the population mean infectiousness $A$, and given these, DLS inversions for the individual infectiousness $I$.  The solutions are the maximum likelihood kernels given one-hundred stochastic trajectories for each of the different kernel combinations (rows {\it top} to {\it bottom}).  No smoothing constraint was used in the GLM regression for these recoveries.  The  DLS inversion for $I$ from the recovered $A$ and $D$ kernels, however, employed damping $\delta_{\rm lsqr}=0.1$ for all examples shown.  

\begin{figure}[t!]
\vskip -2cm
\centerline{\includegraphics[scale=0.55]{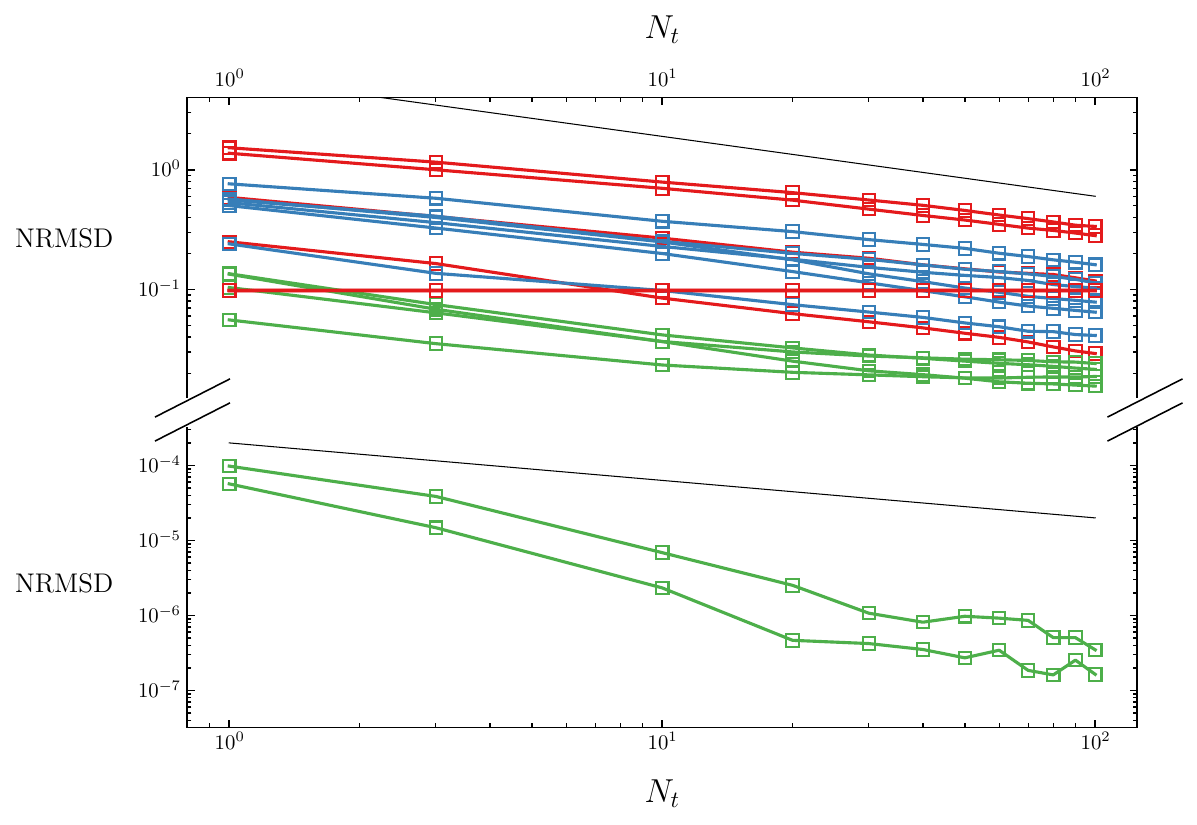}}
\caption{The normalized root-mean-square difference (NRMSD) between the input kernels and those recovered with the unregularlized GLM, $D$ ({\it red}), $B$ ({\it green}), and $A$ ({\it blue}), as a function of the number of trajectories employed $N_t$. {\it Black} fiducial line indicates $N_t^{\,{-0.5}}$ scaling.}
\label{fig:nrmsd}
\end{figure}

With one-hundred trajectories, kernel recovery is quite robust.
The normalized root-mean-square difference (NRMSD, normalized by the range of the kernel) between the input kernels and those recovered with GLM regression ranges from $\sim\!1.6\times 10^{-7}$ to $\sim\!0.34$ (right-hand most points in Figure~\ref{fig:nrmsd}).
In Figure~\ref{fig:nrmsd}, NRMSD is plotted as a function of the number of trajectories $N_t$ employed in the kernel recovery, with each marker indicating the mean value of the NRMSD over ten groups of $N_t$ trajectories.
The fits show consistent improvement as the number of trajectories increases (with one key exception, addressed below), scaling approximately with one over the square-root of the number of trajectories employed (black fiducial line in Figure~\ref{fig:nrmsd}).

In general, the $B$ kernel is recovered more accurately than the other kernels because its recovery depends on the infectious number rather than changes in the population counts with time, as the others do.  
However, the gradient of the recovered $B$ has fluctuations of the same order as those of the recovered $D$, so using it in the recovery of $I$ offers limited advantage. 
Even with DLS regularization, the individual infectiousness kernels recovered show the largest errors, with NRMSD values between $\sim\!0.19$ to $\sim\!0.64$ (not plotted). 

One notable case is the delta-function duration distribution, for which $D$ and $B$ can be recovered almost exactly (last two rows of Figure~\ref{fig:multitrajectoryinversion}).
This is true independent of the number of trajectories used in the inversion and can be understood in light of the fact that for $\delta$-distributed duration the recovery trajectory is a delayed copy of the infection trajectory.
In the two cases shown for which the duration is $\delta$-distributed, NRMSD$\ =0.1$ for all $N_t$ (only one horizontal {\it red} line at NRMSD$\ =0.1$ is apparent in Figure~\ref{fig:nrmsd} for these cases as the two overlie one-another).
This value of the NRMSD reflects a small error in the height of the $\delta$-function distribution recovered.
That height error is nearly independent of the number of trajectories employed, and
the recovered kernels retain their $\delta$-function shape, with values at least
$10^{-8}$ times smaller than the peak for $T \ne1$. 

Despite this ability to recover the distribution of $D$ very precisely when it is  $\delta$-distributed, and along with well recovered $A$ kernels (Figure~\ref{fig:multitrajectoryinversion}), the DLS inversions for the individual infectiousness $I$ struggle in these cases.  The sparsity of the $D$ matrix that results from the $\delta$-distributed infection duration makes it difficult to recover $I$ with the iterative solver 
we employed or other solvers we have tested.  
While it is likely possible to avoid this problem by taking the $\delta$-distributed duration explicitly into account, in practice the functional form of $D$ is unknown and that advantage can not be leveraged. 

\begin{figure}[t!]
\vskip -2.0cm
\centerline{\includegraphics[scale=0.6]{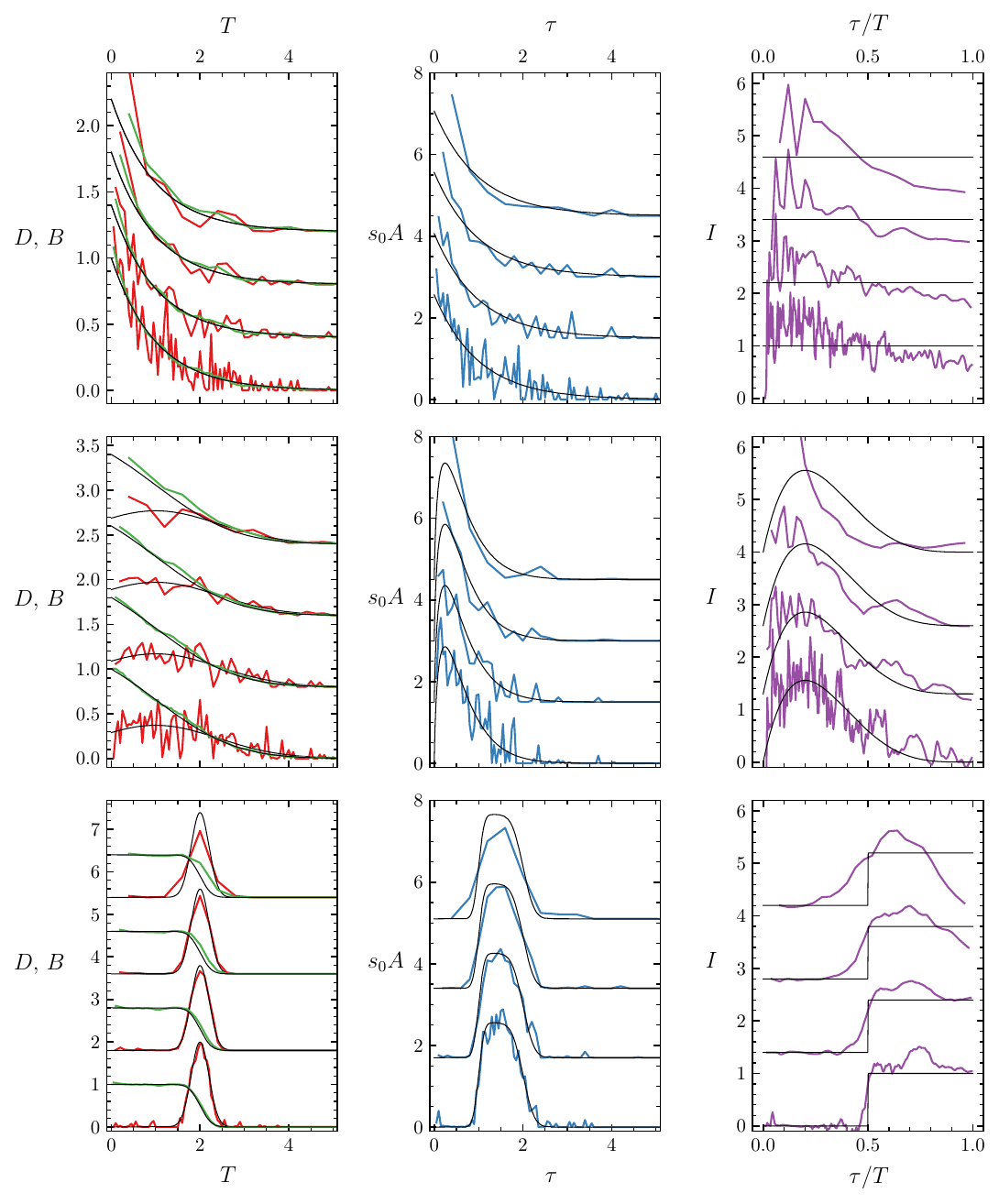}}
\caption{Multi-trajectory Poisson GLM inversions for the disease attribute kernels, $D$ ({\it red}) and $B$ ({\it green}) in the {\it first} column and $s_0A$ ({\it blue}) in the {\it middle} column, along with a regularized least square inversions (using the recovered $D$ and $A$ kernels) for $I$ ({\it purple}) in the {\it third} column.  In each panel, the kernels recovered using trajectory sampling rates $0.05\tau_D$, $0.1\tau_D$, $0.2\tau_D$,  and $0.4\tau_D$ are offset vertically, with coarser sampling solutions uppermost.  One-hundred stochastic trajectories 
(those plotted Figure~\ref{fig:stochastic_trajectories}) were employed in the inversions, each constructed with the target kernels shown in {\it black}. Rows from {\it top} to {\it bottom} for cases EC, N1B, and N2H.}
\label{fig:resolution}
\end{figure}

Inversion dependency on the trajectory sampling interval is illustrated for select kernels in Figure~\ref{fig:resolution}.  Kernel recovery is still quite robust with less frequent sampling of the population trajectories, albeit with some kernel smoothing and an inability to recover kernel structure on time scales shorter than the sampling interval.   While quantitatively the root-mean-square difference between the true kernel and that recovered depends in detail on the particular kernel being considered, typically the difference between the two initially decreases as the sampling interval increases (from $0.05\tau_D$ to $0.1\tau_D$ in Figure~\ref{fig:resolution} for example) because the recovered kernels become smoother before increasing with further increase of the sampling interval (from $0.2\tau_D$ to $0.4\tau_D$ in Figure~\ref{fig:resolution} for example) due to the loss of temporal resolution.  While we focus the remainder of this paper on the highest sampling rate when assessing reporting error and regularization sensitivities, we note that important qualitative disease attributes can be recovered from the population trajectories even with quite infrequent population sampling.

\subsection{Multi-trajectory inversions with observational noise}\label{sec:glm1b}

To add observational error to our trajectories before inverting, we subsample the fully resolved trajectories as previously, compute $\Delta s$ and $\Delta r$ at each time step, and introduce noise at each observational time step (as described in Section~\ref{sec:obs} Equations~(\ref{e1}) and~(\ref{e2})) to obtain $\Delta s_{\rm obs}$ and $\Delta r_{\rm obs}$. We accumulate these observations uncorrected to produce $s_{\rm obs}$ and $i_{\rm obs}$ (as in Equations~(\ref{e3})~-~(\ref{e5})) as needed for the particular kernel inversion.  
For illustration, we consider two cases:  taking the true-positive reporting probability at each time step to be $p_{\textrm{T}}=0.85$ without any false positive reports ($p_{\textrm{F}}=0$) or, alternatively, taking the true-positive reporting probability at each time step to be 
$p_{\textrm{T}}=0.85$ along with a false positive probability per unit time of 
$p_{\textrm{F}}=0.01$.

\begin{figure}[t!]
\vskip -2.5cm
\centerline{\includegraphics[scale=0.45]{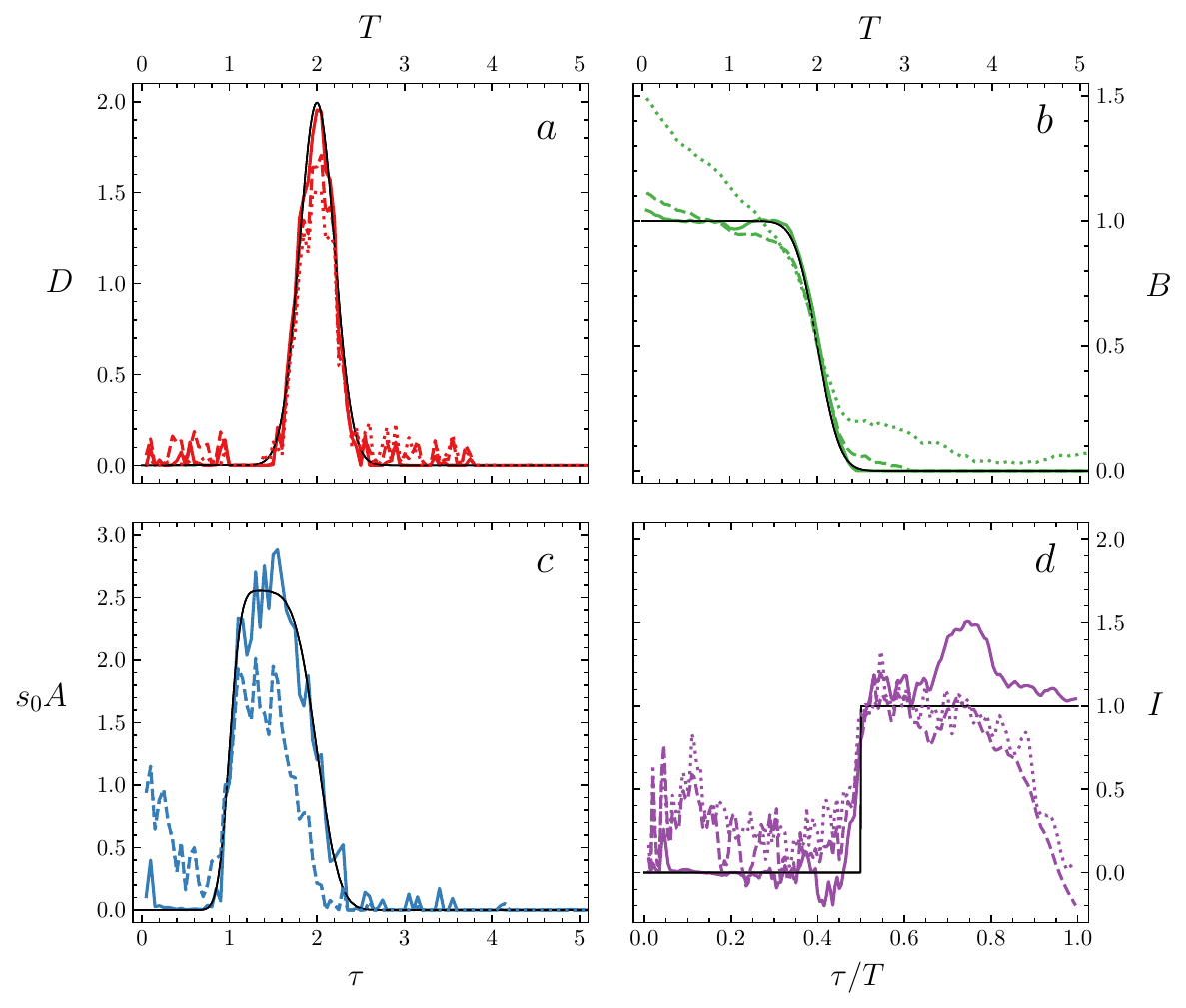}}
\caption{Observational error causes some degradation of disease attribute kernels recovered.  Shown for Case N2H, are example multi-trajectory Poisson GLM inversions for the disease attribute kernels, $D$ ({\it red}), $B$ ({\it green}), and $s_0A$ ({\it blue}) along with a regularized least square inversions for $I$ ({\it purple}).  
Kernels recovered from one-hundred trajectories without observational noise, missing reports, and both missing reports and false positive reports, are plotted with {\it solid}, {\it dashed}, and {\it dotted} line styles respectively.  True kernels used in trajectory  construction are shown in {\it black}.  
}
\label{fig:noise}
\vskip1.0cm
\centerline{\includegraphics[scale=0.45]{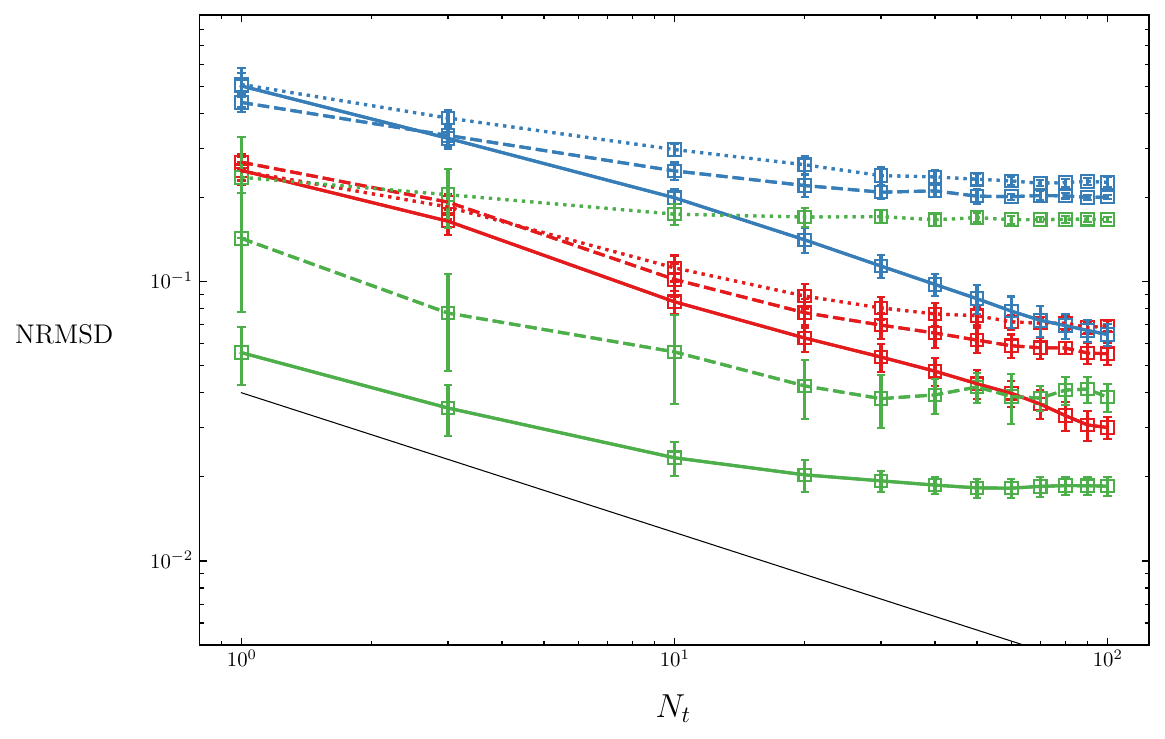}}
\caption{The normalized root-mean-square difference (NRMSD) between the input kernels and those recovered with the unregularlized GLM, $D$ ({\it red}), $B$ ({\it green}), and $A$ ({\it blue}), as deduced from epidemic trajectories without observational noise ({\it solid}), missing reports ({\it dashed}), and both missing reports and false positive reports ({\it dotted}),
as a function of the number of trajectories employed $N_t$.  {\it Black} fiducial line indicates $N_t^{\,{-0.5}}$ scaling.}
\label{fig:nrmsdwithnoise}
\end{figure}

Figure~\ref{fig:noise} plots examples of the resulting GLM kernel fits.    
The inversions were based on one-hundred stochastic trajectories without observational error ({\it solid} curves, as those in Figure~\ref{fig:multitrajectoryinversion} fourth row from top), with incomplete reporting error ({\it dashed} curves, $p_t=0.85$ and $p_f=0$), and with both incomplete reporting and false positive reporting error ({\it dotted} curves, $p_t=0.85$ and $p_f=0.01$).

Figure~\ref{fig:nrmsdwithnoise} displays the normalized root-mean-square difference (NRMSD) between the input kernels and those recovered, $D$ ({\it red}), $B$ ({\it green}), and $A$ ({\it blue}), as deduced from epidemic trajectories without observational noise ({\it solid}), missing reports ({\it dashed}), and both missing reports and false positive reports ({\it dotted}),
as a function of the number of trajectories employed $N_t$ in the regression.  
Each symbol indicates the mean value of the NRMSD over ten groups of $N_t$ trajectories, with error bars of that indicate one standard deviation from that mean. While the variance in NRMSD does tend to decrease with increased number of trajectories, the error bars in the plot are underestimated for $N_t\gtrsim 20$ because the groups of trajectories used are not fully independent.

While kernel recovery remains possible even in presence of quite significant measurement error, the effect of that error is more dramatic for some kernels than others. Recovery of the infection duration distribution $D$ is least sensitive to observational noise, while that of the infection survival kernel $B$, which is extremely well recovered in the presence of stochastic variations alone, is most sensitive to error, particularly false positive reports.  Since $D$ and $B$ are closely related, and either can be used in the determination of $I$, this suggests differing strategies my be favored depending on the anticipated level of observational/reporting error.  Moreover, observational error, causes the NRMSD to plateau, particularly in the case of regressions for $B$ and $A$, with diminishing returns for $N_t\gtrapprox20$ in the regression.

\begin{figure}[t!]
\vskip -2.0cm
\centerline{\includegraphics[scale=0.55]{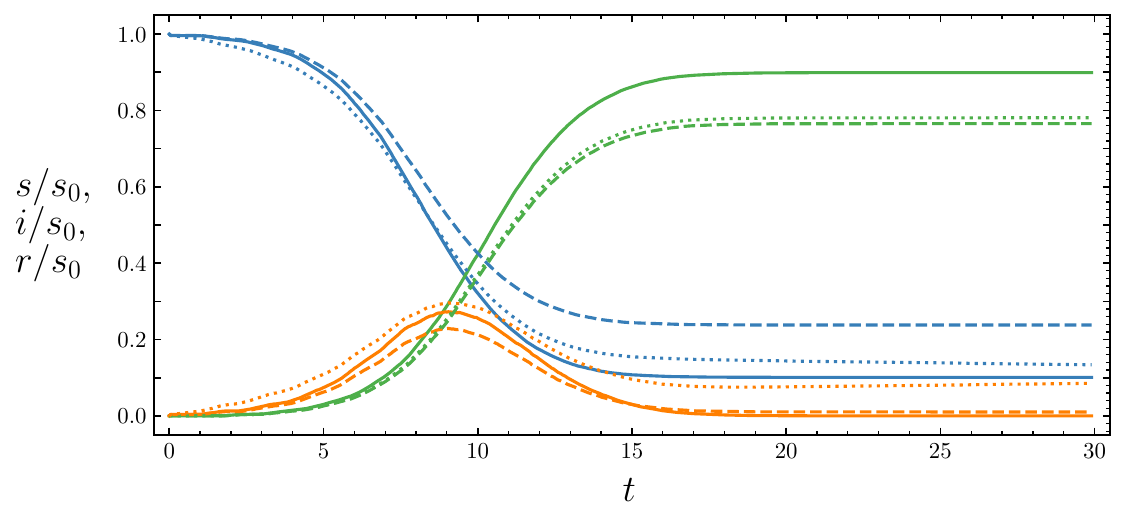}}
\caption{Averages of the epidemic trajectories used in the kernel regressions shown in Figure~\ref{fig:noise}. Population normalized $s$ ({\it blue}), $i$ ({\it orange}), and $r$ ({\it green}),  
without observational noise ({\it solid}), fractional reporting ({\it dashed}), and with both missing reports and false positive reports ({\it dotted)}.
}
\label{fig:avtrajwithnoise}
\end{figure}

These kernel recovery sensitivities are a consequence of the accumulated error in $i_{\rm obs}$ and $s_{\rm obs}$ that results from reporting error in $\Delta s$ and $\Delta r$.
The enhanced error sensitivity of the $B$ reflects the reliance on $i_{\rm obs}$, which suffers from two sources of accumulated of error when $\Delta s_{\rm obs}$ and $\Delta r_{\rm obs}$ are used to obtain $s_{\rm obs}$ and $r_{\rm obs}$  (Equations~(\ref{e3})~$-$~(\ref{e5})).  Regression for the mean infectiousness kernel $A$ is similarly affected by accumulated error, but in $s_{\rm obs}$ alone.
Figure~\ref{fig:avtrajwithnoise} illustrates the systematic changes that occur in the observed epidemic trajectories due to the accumulation of reporting errors, plotting the average of the one-hundred trajectories used to recover the kernels shown in Figure~\ref{fig:noise}.
Underreporting error results in $s_{\rm obs} >s$ and $r_{\rm obs} < r$ over the entire trajectory.  Since $p_{\textrm{T}}$ is taken to be the same for both quantities,  $i_{\rm obs} < i$ at its peak but with nearly full recovery of the population by the end of the outbreak ($i_{\rm obs}$ for $t$ large is $\approx 0$).  With the addition of false positive reporting for both $\Delta s$  and $\Delta r$, both $s_{\rm obs}(t)$
and $r_{\rm obs}(t)$ move closer to their noise free trajectories, but since $p_{\textrm{F}}$ is again chosen to be the same for both quantities $r_{\rm obs}(t)$ remains much further from its true value.  This is because the number of false positive reports of recovery depends on the number infected individuals while the number false positive reports of infection depends on the number of susceptible individuals and $s > i$ over the full trajectory.  The differential error in $s_{\rm obs}(t)$
and $r_{\rm obs}(t)$  leads to a significant number of unrecovered individuals at long times. 

The recovered $A$ and $B$ kernels reflect these systematic changes to the epidemic trajectories as well as the less systematic errors in $\Delta s_{\rm obs}$.  
Importantly, the systematic errors in $i_{\rm obs}(t)$ and $s_{\rm obs}(t)$ produce systematic distortions to the kernels that do not decrease with increasing $N_t$ at high $N_t$ ({\it blue} and {\it green}, {\it dotted} and {\it dashed} curves in Figure~\ref{fig:nrmsdwithnoise}), including  
increased kernel amplitude at low $\tau$ or short $T$ ($A$ and $B$ respectively), the asymmetric
peak of $A$, and the consequent distortion of the deduce individual infectiousness profile $I$ (all in Figure~\ref{fig:noise}).  The multi-trajectory $A$ and $B$ kernels recovered based on smaller numbers of trajectories show the same systematic differences with respect to the input kernels.  In turn, at the error levels we have included, these distortions lead to a $\sim25\%$ reduction in the $R_0$ value deduced by integration of the inverted $A$ kernels . 
Overall the duration distribution $D$ is much less sensitive to observational error as it depends only on $\Delta s_{\rm obs}$ and $\Delta r_{\rm obs}$ at each time step, not on the population trajectories with time.

\subsection{Regularized single-trajectory inversions}\label{sec:glm2}

In the context of an actual epidemic, aggressive monitoring of many individual outbreaks may be challenging or slow. Deductions based on only one or a few well observed population trajectories are therefore invaluable. As illustrated by Figure~\ref{fig:nrmsd}, absent smoothing constraints, the errors in the recovered kernels decrease approximately with the square root of the number of trajectories employed.  Regularization during inversion allows kernel recovery with a smaller number of trajectories or even a single trajectory, albeit with the possibility of some loss in kernel fidelity.
	
\begin{figure}[t!]
\vskip -2cm
\centering
\begin{subfigure}{}
\includegraphics[scale=0.65]{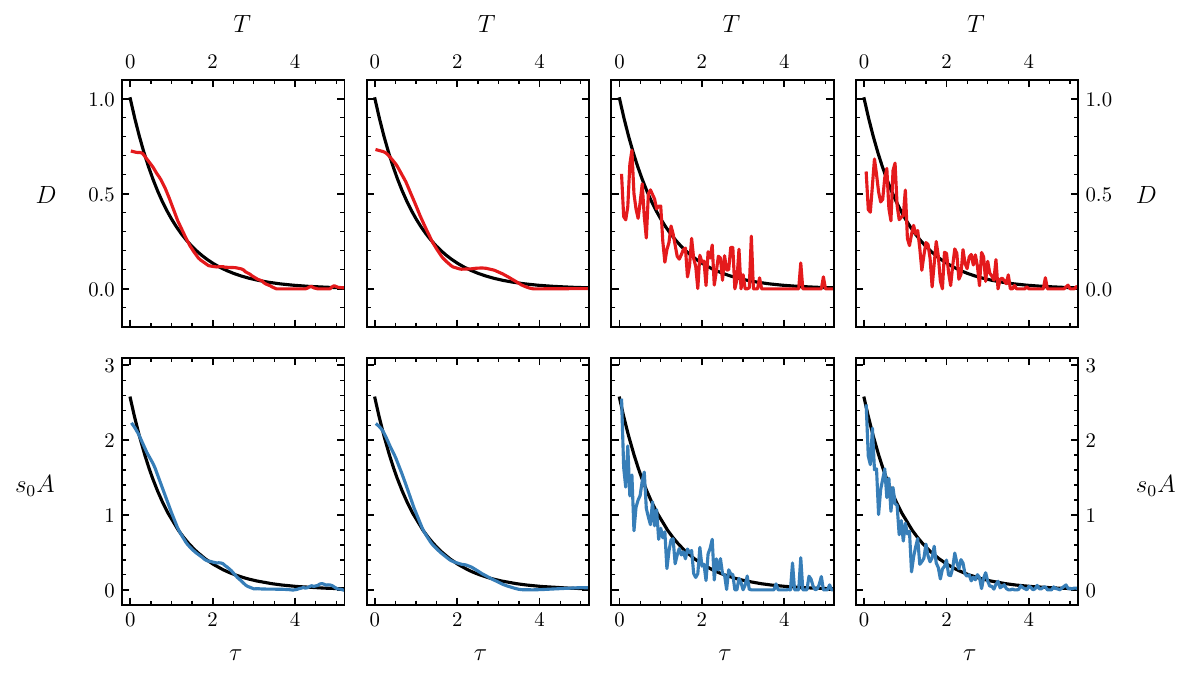}
\end{subfigure}
\vskip0.25cm
\begin{subfigure}{}
\includegraphics[scale=0.65]{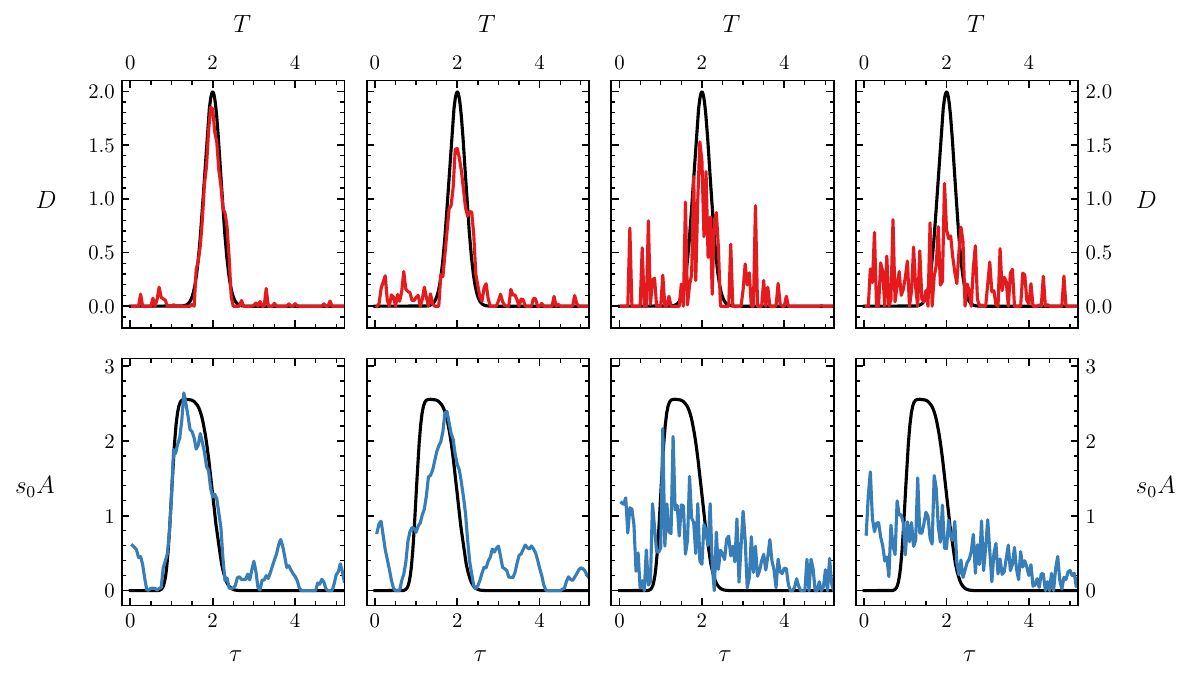}
\end{subfigure}
\caption{Example single trajectory inversions for $D$ and $A$ for epidemic cases EC ({\it top} two rows) and 
N2H ({\it bottom} two rows).  The regressions employed our GLM with different noise and regularization models:  Poisson with adjacent-value smoothing ({\it left most} column), Gaussian with adjacent-value smoothing ({\it second} column),  Poisson with L2 regularization ({\it third} column), Gaussian with L2 regularization ({\it right most} column). 
}
\label{fig:singletraj}
\end{figure}

In Figure~\ref{fig:singletraj} we plot the results of inversions for the disease attribute kernels 
based on a single noise free trajectory. Two different kernel pairs are shown:  $D$ and $A$ for epidemic EC ({\it top} pair) and 
$D$ and $A$ for epidemic N2H ({\it bottom} pair). 

Four inversion variants are illustrated. 
The solutions in the left-hand column employed the Poisson GLM (Poisson negative log-likelihood cost) with adjacent-value smoothing, as described in Section 4.2.  For the solutions in the second column, the noise model was changed to Gaussian (Gaussian negative log-likelihood cost) while maintaining the adjacent-value smoothing.  For the third column, the Poisson GLM was employed with L2 regularization (ridge regression) rather than adjacent-value smoothing, and for the fourth column the noise model was Gaussian with ridge regression rather than adjacent-value smoothing. In all cases we take the value of $\lambda_s$ to be that which yields the minimum NRMSD between the recovered and input kernels.  This allows us to compare `optimally' regularized solutions.  

While the kernel inversions are shown for single trajectories, and different individual trajectories produce somewhat different recovered kernels, the results are typical.  
The Poisson GLM with adjacent-value smoothing recovers the 
kernels most reliably from single trajectories. 
While for some kernels the schemes all do tolerably well (the EC kernels in Figure~\ref{fig:singletraj} for example),
for others both components of our GLM improve kernel recovery reliability (the N2H kernels in Figure~\ref{fig:singletraj} for example).
For best results, the loss function must account for both the Poisson nature of the stochastic noise and favor local smoothness over minimization of the kernel magnitude.  
Moreover, we find that schemes that act as high frequency filters (such the truncated singular value decomposition or Fourier deconvolution (as in \cite{pijpers2021}, not shown here) often fail to recover high spatial-frequency components of the disease attribute kernels.
The population mean infectiousness kernel $A$ in particular can show rapid changes over short post-infection time scales even when the input $D$ and $I$ are quite smooth (see e.g., {\it grey} and {\it gold} curves in Figure~\ref{fig:disease_attributes}$c$).     

We note that, the adjacent-value smoothing we employ also approximately preserves the integral over $A$.  We examined one-hundred single trajectory inversions with fixed $\lambda_s$ for each of the six $A$-kernels we have studied and measured $R_0$ as the integral over the recovered $A$-kernel.  The mean $R_0$ deduced from those inversions is less than 6\% different from its true value in all six cases.  

Nonetheless, regularization can lead to deterioration of the solution.   
An interesting case is that of the $\delta$-distributed duration kernel, which is extremely well recovered without smoothing even from single trajectories (Section~\ref{sec:glm1}).  For that kernel, non-zero $\lambda_s$ systematically broadens the result.  This suggests that a posteriori deduction of the most appropriate smoothing based on systematic changes in the recovered kernel width with $\lambda_s$ may be possible, but how to more generally obtain the optimal smoothing constraint from the data itself, while beyond the scope of this paper, remains an important future research goal.

\section{Conclusion and Implications}\label{sec:imp}

In this paper we examined the ability to determine individual-scale disease attributes from population level epidemic trajectories, probing whether, in the context of a larger epidemic, observing the stochastic evolution of outbreaks in local well-mixed populations allows determination of the underlying disease properties.
We constructed stochastic Kermack-McKendrick trajectories base on specified input kernels and developed a Poisson GLM to recover them.
We found that this approach is effective for a wide range of sampling frequencies, with sample sizes as small as a single trajectory, and in the presence of noise.

Notably, we have shown that epidemic trajectories generated with variable infectivity and/or non-exponential infection duration distributions show important differences from SIR models with the same $R_0$ and the same average infection duration.
These trajectory differences allow, both in the presence of observational error and without such error, the recovery of 
the population mean infectiousness as a function of time since infection $A$, the infected duration distribution $D$, and the infected survival distribution $B$ from the population level stochastic time series. 
Additionally, under the assumption of self-similarity, the individual infectiousness profile $I$ can be 
deduced from the recovered $A$ and $D$ (or $B$) kernels.  

Based on the integro-differential equation formulation of the SIR model, we adopted a natural regression approach to fit the integral kernels.  The two key ingredients in that approach are a proper accounting of the Poisson stochastic noise and regularization based on adjacent value smoothing, when it is needed. 
With these ingredients
we can reliably recover the key epidemic property $A$ based only on observations of  the incidence of new infections, and $D$ and $B$ with the addition of observations of the incidence of recoveries.  We can do so for a variety of kernels, and from the recovered kernels we can disentangle the effects of infectious disease duration $D$ and the variable infectivity $I$.  

If multiple trajectories are employed, the Poisson GLM we developed requires no regularization for kernel recovery, with the accuracy of the inversion scaling roughly as the square root of the number of trajectories employed.  In the presence of observational noise inversion is still possible, but depending its amplitude, systematic distortions can be introduced in the recovered $A$ and $B$ kernels, as they depend on the population trajectories as well as their changes with time and are thus subject to cumulative error.  The $D$ kernel is less susceptible to observational error because its recovery depends only on the changes in the population state at each time step.

With regularization, single trajectory inversions for the disease attributes is also robustly possible.
As the single trajectory regressions perform well,
the approach we have developed is a good fit for the characterization of the 
spread of novel infectious diseases based on limited population monitoring of early outbreaks.

\appendix
\section{Statistical model derivation} \label{sec:appendix:stochastic_model_derivation}

Here we derive our stochastic generative model of an epidemic.
Consider a well-mixed population of size $N$.
Denote the number of susceptible individuals in the population at any time by $s$, and the number of individuals who were first infected $k$ units of time in the past by $n_k$.
Note that this is an alternative way to describe the system of susceptible, infectious, and recovered subpopulations.
The infectious and recovered populations, having all been infected at some point, will be distributed across the bins $n_k$, so $s + \sum_k n_k = N$.
Equivalently, of the  $n_k$ individuals first infected $k$ units of time in the past, some will have recovered, and some will remain infected, so we should expect $n_k$ to contain both type of individuals. 
This notation is equivalent to the notation in the main text if we set $\Delta s_{n-k} = n_k$, but we will use the latter here for clarity. 

If the time between contacts in the population is exponentially distributed with contact rate $r$, then in time $\Delta t$ the number of contacts among individuals in the population is Poisson distributed
\begin{equation}
    p(m \textrm{ contacts}) = \frac{e^{-r\Delta t} }{m!}(r \Delta t)^m.
\end{equation}
Given that there is a contact, we randomly choose two individuals in the population to contact each other (this is the well-mixed assumption), so the probability of contact between a susceptible individual and one of the $n_k$ individuals first infected $k$ units ago is
\begin{equation}
    p(\textrm{contact between members of } s, \ n_k | \textrm{contact}) = 2\frac{s}{N}\frac{n_k}{N}.
\end{equation}
Given contact between a susceptible individual and a previously infected individual, the probability of a new infection occurring is an infection-age dependent quantity
\begin{equation}
    p(\textrm{new infection} | \textrm{contact between members of } s, \ n_k  ) = p_\textrm{infection}(k).
\end{equation}
Marginalizing over $k$, gives the total infection probability given a contact

\begin{equation}
    p(\textrm{new infection} | \textrm{contact}) = \sum_k 2\frac{s}{N} \frac{n_k}{N}  p_\textrm{infection}(k) \label{eq:appendix_infection_prob}.
\end{equation}

If we assume that \textit{each contact event is independent}, then the total number of infections in a time interval with $m$ contacts will be binomially distributed with probability $p(\textrm{new infection} | \textrm{contact})$.
Marginalizing over the (Poisson distributed) total number of contacts in the interval, gives a Poisson distributed number of infections
\begin{equation}
    n_0 \sim \textrm{Pois} \left( \left( r \Delta t \frac{2}{N^2}\right) s \sum_k n_k  p_\textrm{infection}(k) \right)
\end{equation}
Note that the independence assumption \textit{does not} necessarily hold, and renders this an approximation to the ground truth epidemic.  
Independence of contacts within each timestep ignores the facts that 1) each newly infected individual represents an additional source of infections, and 2) each newly infected individual cannot be infected again by another contact in the same time interval.
The approximation becomes exact if we are able to temporally resolve all contact and recovery events, as we do in our simulations, but remains an approximation in statistical fits to coarsely sampled data.

The number of recovered individuals is derived similarly.
Denote the number of individuals who recover in the $i$th time interval by $c_i = \Delta r_i$.
Assume that each individual infection has a duration that is independent of the others, and is chosen from a finite set of time intervals according to
\begin{equation}
    p(\textrm{recovery after $j$ time steps}) = p_\textrm{rec}(j)
\end{equation}
This means that, for example, of the $n_0$ individuals who were infected at time $0$ the number of individuals who recover in each future interval will be multinomially distributed:
\begin{equation}
    c_{0 \cdots m} \sim \textrm{Multi}(n_0, [p_\textrm{rec}(0), p_\textrm{rec}(1), \dots, p_\textrm{rec}(m)])\ ,
\end{equation}
with the expected number of recoveries after $j$ timesteps given by $n_0 p_\textrm{rec}(j)$.
If we now consider all times that individuals were infected, with recoveries being independent between them, then summing over all previous time steps, we have:
\begin{equation}
    \mathbb{E}[c_i] = \sum_k n_{i-k} p_\textrm{rec}(k) \label{eq:appendix_recovery}
\end{equation}
The exact distribution of $c_i$ values is difficult to express in closed form, but if we neglect the correlations between time intervals, then each term in the sum in Eq.~(\ref{eq:appendix_recovery}) is distributed binomially, and therefore the sum can be approximated well by a normal distribution.

From these derivations, we can see that the epidemic kernels in the main text Eqs.~(\ref{eq:KM5},~\ref{eq:KM6}) correspond to:
\begin{align}
    A_k =& \left( r \Delta t \frac{2}{N^2}\right) p_\textrm{infection}(k) \\
    D_k =&\  p_\textrm{rec}(k)
\end{align}
These derivations hold regardless of the forms of $p_\textrm{infection}(k)$ and $p_\textrm{rec}(k)$.
However, not all combinations of these kernels make sense from the prespective of an epidemic.
In particular, we usually think that `recovery' involves no longer being able to transmit infections to others, which implies that there should be some relationship between $p_\textrm{infection}(k)$ and $p_\textrm{rec}(k)$.
This could take the form of
\begin{equation}
    p_\textrm{infection}(k) = p_{\textrm{infection} | \textrm{survive}}(k)  p_\textrm{survive}(k)
\end{equation}
where $p_\textrm{survive}(k) = 1 - \textrm{CDF}_{\textrm{rec}}(k)$.
Thus, enforcing that recovered infections cannot transmit.

That simple form also has some unsatisfactory features: the independence of survival and infectiousness, which implies that an individual could be highly infectious one day, and then fully recovered the next.
In this work, we instead opt to use a characteristic infectiousness time curve $ I\left(\frac{k}{T}\right)$, which describes how infectious an individual is a fraction $\frac{k}{T}$ into their infection.
Thus every individual goes through the same infectiousness time course, but with different speeds.
In this formulation,
\begin{equation}
    p_\textrm{infection}(k) = \int_{k}^\infty I\left(\frac{k}{T}\right) p_\textrm{rec}(T)\,d\,T\, . \label{eq:appendix:inf_convolution}
\end{equation}
Equation~\ref{eq:appendix:inf_convolution} can also be written in a discretized form by quadrature evaluation of the integral~\citep[][]{10.1145/321150.321157}.   After employing a change of variables $x=\tau/T$, noting that the integrand vanishes at the endpoints of the integration interval, and assuming a maximum infectious duration of $T_{\rm max}$ so that $D(t>T_{\rm max})=0$, a composite trapezoidal-rule quadrature evaluation of the integral yields
 \begin{equation}\label{eq:I2}
A(\tau)=N_q\ \tau\ \sum_{k=1}^{N_q-1}\ {1\over{k^2}}\ I\!\left(k\over{N_q}\right)\ D\!\left({{N_q\tau}\over{k}}\right) \ , 
\end{equation}
where $N_q$ is the number of composite-quadrature points.  At discretely sampled times of constant interval, $\tau=n\,T_{\rm max}/N_q$, this can be rewritten as
 \begin{equation}\label{eq:I2_discrete}
A_n=\sum_{k=n}^{N_q-1}\  {{n\,T_{\rm max}}\over{k^2}}\ D\!\left({{n}\over{k}}\,T_{\rm max}\right)I\!\left({k\over{N_q}}\right)\ ,  
\end{equation}
where the lower limit of the sum captures the condition that $D(t>T_{\rm max})=0$.





\section*{Acknowledgments}
The authors thank Stephen Kissler for discussions and encouragement.
This work was partially supported by a faculty appointment at the University of Colorado, Boulder.

\bibliographystyle{elsarticle-harv}
\bibliography{references}

\vfill\eject

\end{document}